
\input harvmac.tex





\let\d\partial

\let\s\sigma

\def\frac#1#2{{\textstyle{#1\over #2}}}

\def\eqalignD#1{
\vcenter{\openup1\jot\halign{
\hfil$\displaystyle{##}$~&
$\displaystyle{##}$\hfil~&
$\displaystyle{##}$\hfil\cr
#1}}
}
\def\eqalignT#1{
\vcenter{\openup1\jot\halign{
\hfil$\displaystyle{##}$~&
$\displaystyle{##}$\hfil~&
$\displaystyle{##}$\hfil~&
$\displaystyle{##}$\hfil\cr
#1}}
}

\def\eqalignQ#1{
\vcenter{\openup1\jot\halign{
\hfil$\displaystyle{##}$~&
$\displaystyle{##}$\hfil~&
$\displaystyle{##}$\hfil~&
$\displaystyle{##}$\hfil~&
$\displaystyle{##}$\hfil\cr
#1}}
}

\def\eqalignS#1{
\vcenter{\openup1\jot\halign{
\hfil$\displaystyle{##}$~&
$\displaystyle{##}$\hfil~&
$\displaystyle{##}$\hfil~&
$\displaystyle{##}$\hfil~&
$\displaystyle{##}$\hfil~&
$\displaystyle{##}$\hfil~&
$\displaystyle{##}$\hfil\cr
#1}}
}

\def\text#1{\quad\hbox{#1}\quad}

\def\la{\lambda}

\def\La{\Lambda}

\def\qb{\bar{q}}

\def\A{{\cal{A}}}
\def\B{{\cal{B }}}
\def\C{{\cal {C}}}

\def\F{{\cal{F}}}

\def\lah{{\hat \lambda}}

\def\y{{\infty}}
\def\vp{{\varphi}}

\def\rw{\rightarrow}

\def\R{\rangle}

\def\su{\widehat{su}}


\newcount\eqnum
\eqnum=0
\def\eq{\eqno(\secsym\the\meqno)\global\advance\meqno by1}
\def\eqlabel#1{{\xdef#1{\secsym\the\meqno}}\eq }

\newwrite\refs
\def\startreferences{
 \immediate\openout\refs=references
 \immediate\write\refs{\baselineskip=14pt \parindent=16pt \parskip=2pt}
}
\startreferences

\refno=0
\def\aref#1{\global\advance\refno by1
 \immediate\write\refs{\noexpand\item{\the\refno.}#1\hfil\par}}
\def\ref#1{\aref{#1}\the\refno}
\def\refname#1{\xdef#1{\the\refno}}

\newcount\exno
\exno=0
\def\Ex{\global\advance\exno by1{\noindent\sl Example \the\exno:

\nobreak\par\nobreak}}

\parskip=6pt

\overfullrule=0mm

\def\frac#1#2{{#1 \over #2}}

\let\d=\partial
\let\La=\Lambda

\def\suh{{\widehat {su}}}
\def\uh{{\widehat u}}
\def\rw{{\rightarrow}}

\newwrite\refs
\def\startreferences{
 \immediate\openout\refs=references
 \immediate\write\refs{\baselineskip=14pt \parindent=16pt \parskip=2pt}
}
\startreferences

\refno=0
\def\aref#1{\global\advance\refno by1
 \immediate\write\refs{\noexpand\item{\the\refno.}#1\hfil\par}}
\def\ref#1{\aref{#1}\the\refno}
\def\refname#1{\xdef#1{\the\refno}}


\Title{\vbox{\baselineskip12pt
\hbox{
}}}
{\vbox {\centerline{ Parafermionic character formulae }}}

\smallskip
\centerline{ P. Jacob and P. Mathieu\foot{Work
supported by NSERC (Canada) and FCAR (Qu\'ebec) }
}

\smallskip\centerline{ \it D\'epartement de
physique,} \smallskip\centerline{Universit\'e Laval,}
\smallskip\centerline{ Qu\'ebec, Canada G1K 7P4}
\smallskip\centerline{(pjacob@phy.ulaval.ca, pmathieu@phy.ulaval.ca)}
\vskip .2in
\bigskip
\bigskip
\centerline{\bf Abstract}
\bigskip
\noindent
We study various aspects of parafermionic theories such as the precise
field content, a
description of a basis of states (that is, the counting of independent
states in a freely generated highest-weight module) and the explicit
expression of the parafermionic singular vectors in
completely irreducible modules.
This analysis culminates in the presentation of new
character formulae for the
$Z_N$ parafermionic primary fields.  These characters provide novel field
theoretical expressions for $\su(2)$ string functions.


\Date{05/00\ }

\let\n\noindent


\newsec{Introduction}

The free-fermion description  of the Ising model is a crucial ingredient in
our rather elaborate understanding of its transition from the conformal point to its
off-critical version.  This property definitely distinguishes it from the other minimal
models.  In fact, this description of the Ising model  makes rather direct the contact with its
statistical formulation. The Ising model is also singled out in that this is the only case for
which the correlation functions have been proved to satisfy (integrable)  differential
equations which reduce, at criticality, to the precise differential equations that follow from
the structure of the singular vectors (see for instance [\ref{B.M. McCoy and J.H.H.
Perk, Nucl. Phys. {\bf B285} [FS19] (1987) 279.}]).

The most natural way of extending these results is probably to look for
conformal theories whose formulation conserves much of the structure of their
statistical relative and which generalizes the Ising model in the most natural way. 
That singles out the three-state Potts model and, more generally, the parafermionic
theories [\ref{A.B. Zamolodchikov and V.A. Fateev. Sov. Phys. JETP {\bf 82} (1985)
215.}\refname\ZFa].

A very simple question that one can ask as a sort of probe of the
smoothness of the transition from the Ising model to generic parafermionic theories is related
to character formulae.  The free-fermionic description of the Ising model  leads to simple
character expressions. These are characters in a free-fermionic Fock space, and these have a
simple decomposition into characters of irreducible Virasoro modules:
$$\eqalign{
& \chi_{1,1} ={q^{-1/48}\over 2}\left [\prod_{n=1}^\y (1+q^{n-1/2})+
\prod_{n=1}^\y (1-q^{n-1/2})\right]\cr
& \chi_{2,1} ={q^{-1/48}\over 2}\left [\prod_{n=1}^\y (1+q^{n-1/2})-
\prod_{n=1}^\y (1-q^{n-1/2})\right]\cr
&\chi_{1,2} =q^{1/24}\prod_{n=1}^\y (1+q^{n})\cr}\eqlabel\isingca$$
However, a crucial aspect of the simplicity of the Ising free-field
representation lies in the fact that this particular representation
captures the whole set of
singular vectors.  It is faithful in that sense.  In other words,
there is no singular-vector
subtraction.  On the other hand,  we do not expect  the parafermionic description to be
faithful.

But in order to construct the character directly
from the parafermionic algebra, some aspects
 of the parafermionic representation
theory are needed.

In contradistinction to the situation with most extended conformal
algebras, the representation theory of the parafermionic algebra has not
been studied in much detail.  Indeed, apart from the
initial fundamental paper of Zamolodchikov
and Fateev [\ZFa], which was followed
by  an  extension dealing mainly with the relation to the
$N=2$ supersymmetric theories [\ref{A.B.
Zamolodchikov and V.A. Fateev, Sov. Phys. JETP  {\bf 63} (1986) 913.}\refname\ZFb],
and a
study of a special example of a more general class of models
[\ref{A.B. Zamolodchikov and V.A. Fateev.  Theor. Math.
Phys. {\bf 71} (1987) 163.}\refname\ZFc], most subsequent
articles deal with the parafermionic theory either via a free-field representation (in
particular, singular vectors and  character formulae have been worked out in [\ref{D.
Nemeschansky, Nucl. Phys. {\bf B363}  (1989) 665.}] from BRST methods using a
representation  introduced in [\ref{D. Nemeschansky, Phys. Lett {\bf B224}  (1989)
121.}\refname\Nem ]  in terms of two free bosons -- see also
[\ref{J. Distler and Z.  Qiu,  Nucl.Phys. {\bf B336} (1990) 533.}]) or use a
correspondence with either WZW models (cf. the classification of modular
invariants in [\ref{D. Gepner and Z. Qiu, Nucl. Phys. {\bf B285} [FS19] (1987)
423.}\refname\GQ]) or special minimal models of W algebras (e.g., cf. the analysis
of the integrable perturbations in [\ref{V.A. Fateev, Int J. Mod Phys. {\bf
A6} (1991) 2109.}]).

  This situation is manifestly due to the
rather complicated structure of the algebra, whose mode formulation
involves  infinite sums.

 Among the basic features that have
never been worked
out directly from the parafermionic algebra and which are needed in order
to work out the character formulae, there are the
counting of the number of independent states (i.e., a description of a
basis of states) for a generic parafermionic highest-weight module,
the explicit form
of the singular vectors associated to the spin fields (the parafermionic
primary fields) and the general structure of the corresponding
irreducible modules.

These precise aspects of the
parafermionic models are addressed here.  We start by clarifying  the
field content (that is, the analogue of the Kac table for the Virasoro minimal models)
of a given model, paying due care to the question of field
identifications and compare it with that predicted by the simplest coset
constructions.  We then work out the explicit form of the spin-field singular
vectors.  These singular vectors are charged and it is shown how to generate
 descendents of appropriate relative charge.  For particular cases (e.g. the
Ising and the Potts models), the singular-vector
 descendents of zero relative charge  can all be described in terms of the
Virasoro modes and the familiar Virasoro singular-vector expressions are
recovered; this is  shown in appendix B.   Finally, we present the
parafermionic character formulae for the charged and the uncharged modules built on
the highest-weight states corresponding to the spin fields.

The development of the representation theory of the parafermionic conformal
field theory, which will always refer here to the basic models of [\ZFa], has interesting
offshots. Indeed, many extensions of this class of models can be constructed and some (if
not most) of which may not have a coset construction nor any simple relation to other extended
conformal field theories. Relying upon genuine parafermionic constructions would then seems to
be the only available tool at hand.


\newsec{Parafermionic conformal field theories }

In this section, we review briefly the main results of [\ZFa] that will
be needed in the sequel.

The parafermionic algebra is constructed out of $N$ parafermionic fields
$\psi_k$, $k=0,1,\cdots , N-1$ (with $\psi_0=I$ and $\psi^\dagger_k=
\psi_{N-k}$) of holomorphic conformal dimension
$$h_{\psi_k}={k(N-k) \over N}  \eq$$
that satisfy the OPE:
$$\eqalign{
\psi_k (z) \,\psi_{k'} (w) &\sim c_{k,k'} (z-w)^{-2kk'/N}
 \psi_{k+k'} (w)  \qquad (k'\not=N- k) \cr
\psi_k (z) \,\psi^\dagger_k (w) &\sim (z-w)^{-2k(N-k)/N}
[I+(z-w)^2 {2 h_{\psi_k} \over c}\, T(w)]\cr
T(z) \,\psi_k (w) &\sim {h_{\psi_k} \psi_k (w)
\over (z-w)^2} + {\partial \psi_k (w) \over (z-w)}\cr
T(z) \,T (w) &\sim {c/2\over (z-w)^4}+{2T(w)
\over (z-w)^2} + {\partial T (w) \over (z-w)}\cr}\eq$$
The associativity conditions fix the constant $c_{k,k'}$ to be
$$c^2_{k,k'}={\Gamma (k+k'+1) \Gamma (N-k+1) \Gamma (N-k'+1) \over
\Gamma (k+1) \Gamma (k'+1) \Gamma (N-k-k'+1) \Gamma (N+1)} \eq$$
and the  central charge is linked to $N$ by
$$c={2(N-1)\over (N+2)}\eq$$

The space of fields $\F$
can be decomposed into a direct sum of charged spaces $\F_{[q,{\bar q}]}$, where
$q,{\bar q}$ represent the holomorphic and antiholomorphic charges, defined modulo $2N$.
The charge  normalization is fixed by the charge assignment of the parafermionic fields:
$\psi_k\in \F_{[2k,0]}$ and ${\bar \psi}_k\in \F_{[0,2k]}$.

Let us indicate how this space of fields can be organized in terms of
the parafermionic algebra as defined by the above OPEs.  For this, it is
convenient to describe the action of the parafermionic fields in terms of
their modes.  However,  the precise form of the mode expansion
depends upon the field on which the parafermion acts upon.  For instance,
on a field $\phi_{[q,\qb]}$, the mode expansion of $\psi_1$ and
$\psi_1^\dagger$ is defined as
$$\eqalign{ \psi_1(z)\phi_{[q,\qb]}(0) & = \sum_{m=-\y}^\y
z^{-q/N-m-1}A_{(1+q)/N+m}\, \phi_{[q,\qb]}(0)\cr
\psi_1^\dagger (z)\phi_{[q,\qb]}(0) & = \sum_{m=-\y}^\y
z^{q/N-m-1}A^\dagger_{(1-q)/N+m}\,\phi_{[q,\qb]}(0)\cr}\eqlabel\modep$$
As usual, we read off the dimension of a mode directly from the negative
of its index:
$$\eqalign{ &h_{A_{(1+q)/N+m}\phi_{[q,\qb]}}= -{1+q\over N} -m+
h_{\phi_{[q,\qb]}}\cr
&h_{A^\dagger_{(1-q)/N+m}\phi_{[q,\qb]}}= -{1-q\over N} -m+
h_{\phi_{[q,\qb]}}\cr}\eq$$
and the charge of a mode is that of its corresponding field:
$$ q_{A^{\phantom\dagger}_{(1+q)/N+m}\phi_{[q,\qb]}}= 2+q\, , \qquad \qquad
q_{A^\dagger_{(1-q)/N+m}\phi_{[q,\qb]}}= -2+q\eq$$


 Now, it turns out that the action of $\psi_1$ and $\psi_1^\dagger$
suffices to generate the whole space of fields.   Before we make this
explicit, we display the  mode commutation relation. The action of the modes $A_u$ or
$A^\dagger_v$ on a field with given charge is fixed by the
 inverted version of (\modep)
 $$\eqalign{ &A_{(1+q)/N+m} \phi_{[q,\qb]} (0) = {1\over 2 \pi i}\oint_0 dz\,
  \psi_1 (z)\, z^{q/N+m}\, \phi_{[q,\qb]} (0)
 \cr
 & A^\dagger_{(1-q)/N+m} \phi_{[q,\qb]} (0) ={1 \over 2 \pi i} \oint_0
{dz }\, \psi^\dagger_1 (z)\, z^{-q/N+m}\,
\phi_{[q,\qb]}(0) \cr}\eq$$
The commutation relation between
$A_u$ and $A_v^\dagger$ follows from the computation of the integral [\ZFa]:
$${1\over (2 \pi i)^2}\oint_0 dw\,\oint_w dz\,
 \psi_1 (z)\, \psi_1^\dagger (w)\, z^{q/N+m}\, w^{-q/N+m+1}\,(z-w)^{-(N+2)/N}\,
\phi_{[q,\qb]} (0)\eq$$
and this yields\foot{This are two misprints in [\ZFa] with regard to this expression:
a power of $w$ is missing in the previous integral and the relative
sign in the following commutation relation is incorrect.}
$$\eqalign{ &\sum_{l=0}^{\infty} c_l \left[ A_{-(1-q)/N+n-l-1}
A^\dagger_{(1-q)/N+m+l+1} + A^\dagger_{-(q+1)/N+m-l}A_{(1+q)/N+n+l} \right] \phi_{[q,\qb]}
(0) \cr  &=\left[ {(N+2) \over N} L_{n+m} + {1 \over 2}
(n+{q\over N})(n-1+{q\over N})
\delta_{n+m,0} \right] \phi_{[q,\qb]}(0)\cr}
\eqlabel\comma$$
where
$$c_l= C^{(l)}_{-(N+2)/N}\, , \qquad \quad  C_n^{(p)}= {\Gamma(p-n)\over
p!\Gamma(-n)}\eq$$ To get the commutation relation between $A_u$ and $A_v$,
we need to evaluate
$${1\over (2 \pi i)^2}\oint_0 dw\,\oint_w dz\,
 \psi_1 (z)\, \psi_1 (w)\, z^{q/N+m}\, w^{q/N+m}\,(z-w)^{2/N}\,
\phi_{[q,\qb]} (0)\eq$$
and that leads to
$$\sum_{l=0}^{\infty} c'_l
\left[ A_{(3+q)/N+n-l}A_{(1+q)/N+m+l}-A_{(3+q)/N+m-l}A_{(1+q)/N+n+l} \right]
\phi_{[q,\qb]}(0)=0
\eqlabel\commb$$
where $$c'_l = C^{(l)}_{2/N}\eq$$
There is a similar expression with $A$ and $q$  replaced by $A^\dagger$ and
$-q$. On the other hand, the commutation relations between the Virasoro and the
parafermionic modes are
$$\eqalign{
& [L_m,A^\dagger_u]= -(u+m/N)A_{u+m}^\dagger\cr
& [L_m,A_v]= -(v+m/N)A_{v+m}\cr}\eq$$

The parafermionic primary fields are called the spin fields $\s_k\in
\F_{[k,k]}$, $k=0, 1, \cdots, N-1$.  They are associated to the highest-weight states
$|\s_k\R$ satisfying the  holomorphic highest-weight conditions
$$A_{(1+k)/N+n} | \sigma_k
\rangle = A^\dagger_{(1-k)/N+n+1} | \sigma_k \rangle
=L_{n+1}| \sigma_k \rangle =0  \qquad {\rm for}\quad n\geq 0\eq$$
Actually, it follows from the commutation relation (\comma) with $n=m=0$ that there
can exist only one primary field of a given charge and its conformal dimension is
fixed directly by the algebra to be:
$$h_{\sigma_k}={k(N-k) \over 2N(N+2)} \eq$$
Equivalently, the primary fields could be chosen to be the disorder fields
$\mu_k\in \F_{[-k,k]}$;
the
corresponding states satisfy the holomorphic highest-weight conditions:
$$A_{(1-k)/N+n+1} | \mu_k
\rangle = A^\dagger_{(1+k)/N+n} | \mu_k \rangle
=L_{n+1}| \mu_k \rangle =0  \qquad {\rm for}\quad n\geq 0\eqlabel\dishw$$
and it follows from (\comma) with $n=-m=1$ that $h_{\mu_k}= h_{\sigma_k}$. Notice that
with this definition for the disorder field, we have $\s_k \mu_{-k}\sim
\psi_k$ and
$\s_k\mu_k\sim {\bar \psi}_k$.

The whole set of parafermionic states can be generated from $|\s_k\R$ by the
action of the modes $A_u$  and $A^\dagger_v$.  Since the fractional
part of the modes is fixed unambiguously by the charge of the field or the
state on which it acts, it will be omitted in the following in order to lighten the
notation.  To emphasis the fact that the fractional part has been omitted,
we will write
$\A_n$ for $A_{n+(1+q)/N}$ and $\B_m$ for $A_{m+(1-q)/N}^\dagger$; more
precisely, we have:
$$\A_n |\phi_{[q,\qb]}\R\equiv  A_{n+(1+q)/N}|\phi_{[q,\qb]}\R\, ,\qquad
\B_n |\phi_{[q,\qb]}\R\equiv  A^\dagger_{n+(1-q)/N}|\phi_{[q,\qb]}\R\eq$$
The commutation relations take the simplified forms
$$\eqalign{ &\sum_{l=0}^{\infty} c_l \left[ \A_{n-l-1}
\B_{m+l+1} + \B_{m-l}\A_{n+l} \right] \phi_{[q,\qb]}
(0) \cr  &=\left[ {(N+2) \over N} L_{n+m} + {1 \over 2}
(n+{q\over N})(n-1+{q\over N})
\delta_{n+m,0} \right] \phi_{[q,\qb]}(0)\cr}
\eq$$
and
$$\sum_{l=0}^{\infty} c'_l
\left[ \A_{n-l}\A_{m+l}-\A_{m-l}\A_{n+l} \right]
\phi_{[q,\qb]}(0) =0
\eq$$

{}From $ |\s_k\R$, we can generate charged sectors
by the multiple action
of
$\A_{-1}$ or $\B_0$.  For simplicity, we will focus on the holomorphic sector and write
$$\s_k= \varphi_k^{(0)}{\bar \varphi}_k^{(0)}\eq$$ The repeated action of
the $\A_{-1}$ operators on $| \varphi_k^{(0)}\R$ produces the states
$$| \varphi_k^{(\ell)}\R = \A_{-1}^\ell| \varphi_k^{(0)}\R  \qquad 0\leq \ell\leq
N-k\eqlabel\bb$$ ($\A_{-1}^\ell\equiv (\A_{-1})^\ell$), with charge
$ 2\ell+k$ and dimension $$
h_{\varphi_k^{(\ell)}}= h_{\s_k}+{\ell(N-k-\ell)\over N}\eq$$By acting
with various powers of
$\B_0$ on $| \varphi_k^{(0)}\R$, one gets
$$|\varphi_k^{(-\ell)}\R = \B_{0}^\ell|\varphi_k^{(0)}\R \qquad 0\leq\ell \leq
k\eqlabel\bc$$ of charge
$ -2\ell +k$  and dimension
$$
h_{\varphi_k^{(-\ell)}}= h_{\s_k}+{\ell(k-\ell)\over N}\eq$$
(The bounds on $\ell$ are explained below.)
In particular, we have
$$\psi_k= \varphi_0^{(2k)} \,\quad   {\rm and } \quad \qquad \mu_k=
\varphi_k^{(-k)}{\bar \varphi}_k^{(0)}\eq$$ The parafermionic field $\psi_k$
can thus be obtained from $k$
applications of the generator $\A_{-1}$ on the identity field. This is the announced
result that the different  $\psi_k$ can all be obtained from $\psi_1$. On the other
hand, notice that the disorder fields
$\mu_k$ appear themselves as descendents of the spin fields.  We will see in the next
section how this can be reconciliated with their primary character.
Note that the states $|\varphi_k^{(\pm\ell)}\R $ are
necessarily Virasoro highest-weight states.  It should
be recalled that $\varphi^{(\pm\ell)}_k$ has holomorphic charge $\pm 2\ell+k$.


\newsec{Field content and field identifications}

Let us first comment on the upper bounds for the value of $\ell$ in (\bb) and (\bc):
these limits are fixed by the existence of singular vectors in the cases where
$\ell=N-k+1$ and $\ell=k+1$ respectively.  More precisely, $|\s_k\R$ has the
following two singular vectors (the proof of this statement is reported to section
5):
$$ \A_{-1}^{N-k+1}|\s_k\R =0\, ,\qquad
\B_{0}^{k+1}|\s_k\R=0\eqlabel\siga$$
to which correspond the null fields
 $$\varphi^{(N-k+1)}_k=0\, ,\qquad  \varphi^{(-k-1)}_k=0\eq$$
Therefore the two states at the extremity of the string
$|\vp_{k}^{(\ell)}\R$ satisfy
 $$\A_{-1}|\varphi^{(N-k)}_k\R=0 \, ,\qquad \B_0|\varphi^{(-k)}_k\R=0\eq$$
Rewritting
this with the fractional part of the modes reinserted, this is nothing but
$$\A_{1+(1-k)/N}|\varphi^{(N-k)}_k\R=0 \, ,\qquad
\B_{(1+k)/N}|\varphi^{(-k)}_k\R=0\eqlabel\smu$$
Now, the two fields
$\varphi^{(N-k)}_k$ and $\varphi^{(-k)}_k$
have the same dimension and their charge differ by $2N$.  It has already been indicated
that the charge is defined modulo
$2N$. The parafermionic theory under consideration is essentially an affine $\suh(2)$
theory, up to the description of the charge (this statement is made precise
below, in terms of a coset) and for such theories, the equality of the conformal
dimension ensures the field identification.  Therefore, the fields
$\varphi^{(N-k)}_k$ and $\varphi^{(-k)}_k$ are identical:
$$\varphi^{(N-k)}_k=\varphi^{(-k)}_k\eqlabel\fid$$
Since $\varphi^{(-k)}_k= \mu_k$, these are two representations of the disorder
field.  The singular-vector expressions (\smu) are thus simply
$$\A_{1+(1-k)/N} |\mu_k\R = \B_{(1+k)/N}|\mu_k\R=0\eqlabel\adad$$
The other positive modes annihilate $|\mu_k\R$ (this is made
explicit in section 5).  As a result, the singular-vector conditions are nothing but
the highest-weight conditions (\dishw) on the disorder states.

{}From an heuristic point of view, one can see the existence of these singular vectors
as ensuring the unitarity of the theory, that is, the absence of fields with negative
conformal dimension.  Indeed, observe that by writting explicitly the fractional modes
in
$\A_{-1}^{N-k+1}|\s_k\R$, we get
$$A_{1+(1-k)/N}\, \big[A_{1-(1+k)/N} \cdots A_{-1+(1+k)/N}\big]\, |\s_k\R\eq$$
and the sum of the fractional modes of the $N-k$ rightmost factors add up to zero
(which is also clear from the fact that  $\A_{-1}^{N-k}|\s_k\R=|\mu_k\R $ and that
$h_{\s_k}=h_{\mu_k}$). The dimension of the extra mode is such that when added up to
that of $\s_k$, the sum is negative. In other words, in the string
$\A_{-1}^\ell|\s_k\R$, as $\ell$ increases, the dimension starts by increasing
but at some point this process is reversed and it then decreases continuously; however, the
very first state that acquires a negative dimension is
singular.\foot{If the $\A$--string contains indices less than $-1$, more $\A_{-1}$
operators are needed in order to reach a negative dimension and these additional
operators ensure that the resulting string can be described as a descendent of the
singular vector.} The same remark applies to the
$\B_0^\ell |\s_k\R$ string.

In the holomorphic sector, $\mu_k$ is identical to $\s_{N-k}$: they have the same
dimension and satisfiy the same highest-weight conditions.  
More precisely, $\varphi_{N-k}^{(0)}= \varphi_{k}^{(-k)}$.  Therefore, by including all
the spin primary states and their descendents, we count twice too many states.

To avoid this double counting, we can thus restrict the space of 
fields to those obtained
from
$|\varphi_{k}^{(0)}\R$ and their descendents, with $k=0,\cdots, [N/2]$, (the square bracket
indicates the integer part) when $N$ is odd and include half of the descendents of
$|\varphi_{N/2}^{(0)}\R$ when $N$ is even.  More simply, we could avoid considering
both the $\A$--type  and the $\B$--type descendents and keep only one of these two
sets.  In that case, we should not keep the `holomorphic disorder fields'
 as `holomorphic spin-field' descendents since they already appear in disguised form as
$\vp_{N-k}^{(0)}$.  For instance, if we keep only the
$\A$ descendents, then we need to keep all the $\A$ descendents of $|\vp_{k}^{(0)}\R$
up to the next to
last one (since the last one is the holomorphic restriction of the disorder state,
$|\vp_{k}^{(-k)}\R$).  The resulting set of
independent states $|\vp_k^{(\ell)}\R$ is thus:
$$ \big\{ \,|\vp_k^{(\ell)}\R= \A_{-1}^\ell\, |\varphi_{k}^{(0)}\R \, |\,\, k=0,\cdots N-1\,, \quad 0\leq
\ell
\leq N-k-1\,\big\}\eq$$
The number of such states is simply
$$ \sum_{k=0}^{N-1}\sum_{\ell=0}^{N-1-k} 1= {N(N+1)\over 2}\eq$$

This counting of states can be checked by considering coset realizations of this
parafermionic model. The simplest one is
$\suh(2)_N/\uh(1)$.  The number of independent primary coset fields is also $N(N+1)/2$
(cf. [\ref{P. Di Francesco, P. Mathieu and D. S\'en\'echal, {\it Conformal Field theory},
Springer Verlag, 1997.}\refname\DMS] sect.18.5).  Another one is
furnished by the diagonal coset
$\suh(N)_1\oplus
\suh(N)_1/\suh(N)_2$.  The number of distinct holomorphic WZW primary fileds for the
$\suh(N)_k$ model is $(k+N-1)!/k!(N-1)!$. The branching condition in the coset is simply
taken into account by ignoring one $\suh(N)_1$ factor (there is a unique field associated
to this WZW term that is compatible with the branching condition of the three
coset component-fields). The total number of coset fields must be reduced by a factor $N$
to take care of the  field identifications resulting from  outer
automorphisms (and there are no fixed points here). The number of distinct coset fields is
thus:
$${N!\over (N-1)!}\times{(N+1)!\over 2!(N-1)!}\times{1\over N}= {N(N+1)\over2}\eq$$

Let us list the independent holomorphic fields $\vp_k^{(\ell)}$ for the first few
values of $N$.  For $N=2$, which is the Ising model, we have the following correspondence
with Virasoro primary fields
$$\eqalignD{
& \vp_0^{(0)}= I=\phi_{(1,1)} \qquad\qquad &\vp_0^{(1)}=\psi=\phi_{(2,1)} \cr
& (h=0)   &(h=1/2)\cr
\noalign{\medskip}
& \vp_1^{(0)}= \s= \phi_{(1,2)} &\quad\quad ~\cr
& (h=1/16)&~\cr}\eq$$
(where here and below the spin-field notation refers to its holomorphic restriction).
The $N=3$ parafermionic theory corresponds to
 the three-state Potts model (whose Virasoro partition function has a  non-diagonal
form):
$$\eqalignT{
& \vp_0^{(0)}= I=\phi_{(1,1)}+\phi_{(4,1)} \quad\quad
&\vp_0^{(1)}=\psi_1= \phi_{(1,3)}\quad\quad &\vp_0^{(2)}=\psi_2=\phi_{(1,3)}^*\cr &
(h=0)&    (h=2/3) &(h=2/3)\cr
\noalign{\medskip}
& \vp_1^{(0)}= \s_1=\phi_{(2,3)} \quad\quad
&\vp_1^{(1)}=\epsilon_1=\phi_{(2,1)}+\phi_{(3,1)}\quad\quad &~\cr & (h=1/15)&    (h=2/5)
&~\cr
\noalign{\medskip}
&\vp_2^{(0)}=\s_2=\phi_{(2,3)}^* &~&~\cr
 & (h=1/15)&    ~ &~\cr}\eq$$
The $N=4$ model is an orbifold $c=1$ theory on a rational circle of radius
$R=\sqrt{6}$, with field content:
$$\eqalignQ{
& \vp_0^{(0)}= I \qquad\qquad &\vp_0^{(1)}=\Phi_3^{(1)}\qquad\qquad
&\vp_0^{(2)}=\Theta \qquad\qquad &\vp_0^{(3)}=\Phi_3^{(2)}\cr
& (h=0)&    (h=3/4) &(h=1)&(h=3/4)\cr
\noalign{\medskip}
& \vp_1^{(0)}= \s^{(1)} \qquad\qquad &\vp_1^{(1)}=\tau^{(1)}\qquad\qquad &
\vp_1^{(2)}=\tau^{(2)} &~\cr
& (h=1/16)&    (h=9/16) & (h=9/16)&~\cr
\noalign{\medskip}
&\vp_2^{(0)}=\Phi_1 &\vp_2^{(1)}=\Phi_2& ~&~\cr
 & (h=1/12)& (h=1/3)&   ~ &~\cr
\noalign{\medskip}
&\vp_3^{(0)}= \s^{(2)} &~& ~&~\cr
 & (h=1/16)&~&   ~ &~\cr}\eq$$
(compare with [\ref{R. Dijkgraaf, C. Vafa,  E. Verlinde and H. Verlinde, Comm. Math.
Phys. {\bf 123} (1989) 485.}] and [\DMS] sect. 17.B.6). Here $\s^{(i)}$ and $\tau^{(i)}$
($i=1,2$) are the twist fields of the orbifold theory and $\Theta $ is a dimension-one
field.  The rest of the spectrum depends upon the value of the radius; with $R=\sqrt{2p'}$,
there are the fields $\Phi_\la$ with
$\la=1,2,\cdots, p'-1$ of dimension $\la^2/4p'$ and the doubly degenerate field
$\Phi_{p'}^{(i)}$ of dimension $p'/4$.

\newsec{A basis of states}

We now determine a convenient basis of
states for the description of a generic highest-weight module in the parafermionic
algebra, level by level. The highest-weight state $|\phi_q\R$
is defined
by the condition
$$\A_{-n-1}|\phi_q\R =  \B_{-n}|\phi_q\R  = L_{-n-1}|\phi_q\R=0\quad{\rm for
}\quad n\geq 0\eq$$ As before, we ignore the fractional part of the modes. The results
are presented in the form of lemma, illustrated with simple examples.  The proof of
the first three lemma  is reported in Appendix
A. For the other ones, the proof is obvious.

\n {\it Lemma} 1 -- On a generic highest-weight state $|\phi_q\R$, any state at level
$s$ built out of
$j$ operators
$\A$ can be expressed as a linear combination of terms of the form
$$\A_{-n_1}\A_{-n_2} .. .  \A_{-n_j} |\phi_q\R\eq$$
where $\sum_i n_i=s\geq j$ and
$n_i \geq 1 $ and $n_k \geq n_l$ if $ k < l $.

\n {\it Lemma} 2 -- On a generic highest-weight state $|\phi_q\R$, any state at level
$s$
built out of
$j$ operators
$\B$ can be expressed as a linear combination of terms of the form
$$\B_{-m_1}\B_{-m_2} .. .  \B_{-m_j} |\phi_q\R\eq$$
where $\sum_i m_i=s$ and
$m_i \geq 0 $ and $m_k \geq m_l$ if $ k < l $.

Consider now mixed strings, that is, sequences containing both types of operators. The
rearrangement of such strings will generate $L_m$ terms.  These are easily taken
care of by noticing the relation (that follows from (\comma) with $n=0$):
$$L_m|\phi_q\R\ = {N\over N+2} \sum_{l=0}^\y c'_l\A_{-l-1}\B_{m+l+1}
|\phi_q\R\eqlabel\lla$$ Hence, the Virasoro modes can be replaced by appropriate
sums of
$\A\B$-type terms and can thus be ignored in the counting of independent states.

\n {\it Lemma} 3 -- On a generic highest-weight state $|\phi_q\R$, a state at level
$s$ containing
$j$ operators $\A$ and $k$ operators $\B$ in any ordering, can be written as a linear
combination of terms of the type
$$\A_{-n_1}\A_{-n_2} .. .  \A_{-n_{j'}}\B_{-m_1}\B_{-m_2} .. .  \B_{-m_{k'}}
|\phi_q\R\eq$$
with $$j-k=j'-k'\, , \qquad \qquad \sum_{i=1}^{j'} n_i+ \sum_{i=1}^{k'}
m_i=s\eqlabel\mixst$$ and
$$ n_i \geq 1\qquad n_p \geq n_l\quad {\rm  if}\quad  p < l
\quad {\rm and} \quad m_i \geq 0\qquad m_p \geq m_l\quad {\rm  if}\quad  p < l\eq$$

The main point of this lemma is that we can order separately the $\A$ strings and the
$\B$ strings but there is no mixed ordering.

We illustrate these results with simple examples.  Consider first
$\A_2\A_{-4}  |\phi_q\R$. $\A_2\A_{-4}$ is the first term of the infinite
sum $\sum_{l\geq 0} c_l\A_{n-l}\A_{m+l}$ with $n=2$ and $m=-4$, since $c_0=1$.  All
but a finite number of terms in this sum vanish when it is acted on a
highest-weight state. In fact, we have:
$$\A_2\A_{-4}  |\phi_q\R = \big( \sum_{l=0}^\y
c_l\A_{2-l}\A_{-4+l}-c_1\A_1\A_{-3}
-c_2\A_0\A_{-2}
-c_3\A_{-1}\A_{-1}\big)|\phi_q\R\eq$$
 Using the commutation relation (\commb), we can rewrite the infinite sum
differently, as:
$$\eqalign{ \A_2\A_{-4}  |\phi_q\R &= \big( \sum_{l=0}^\y
c_l\A_{-4-l}\A_{2+l}-c_1\A_1\A_{-3}
-c_2\A_0\A_{-2}
-c_3\A_{-1}\A_{-1}\big)|\phi_q\R\cr
&=  (-c_1\A_1\A_{-3}
-c_2\A_0\A_{-2}
-c_3\A_{-1}\A_{-1}\big)|\phi_q\R\cr}\eq$$
Similarly, $\A_1\A_{-3}|\phi_q\R$ can be expressed as a linear combination of
$\A_0\A_{-2}|\phi_q\R$ and $\A_{-1}\A_{-1}|\phi_q\R$ and finally,
$\A_0\A_{-2}|\phi_q\R$ is found to be proportional to $\A_{-1}\A_{-1}|\phi_q\R$.
Hence, $\A_2\A_{-4}  |\phi_q\R$ is simply proportional to
$\A_{-1}\A_{-1}|\phi_q\R$, as expected, since the latter is the only independent
state at level 2 with relative charge 4.

Consider next $\B_{-3}\A_{-1}|\phi_q\R$; since $c'_0=1$, we can write
$$\eqalign{
 \B_{-3}\A_{-1}|\phi_q\R &=\sum_{l=0}^\y
 c'_l\, \B_{-3-l}\A_{-1+l}|\phi_q\R \cr
&= \left [-\sum_{l=0}^\y
c'_l\, \A_{-2-l}\B_{-2+l}+{N+2\over N}\, L_{-4} \right] |\phi_q\R\cr
&= \left [ -\A_{-2}\B_{-2}-c'_1\A_{-3}\B_{-1} -c'_2\A_{-4}\B_{0} +
{N+2\over N}\,  L_{-4}\right] |\phi_q\R\cr}\eq$$
And using the expression (\lla):
$${N+2\over N}\,  L_{-4}|\phi_q\R = \left [\A_{-1}\B_{-3}+
c'_1\A_{-2}\B_{-2}+c'_2\A_{-3}\B_{-1} +c'_3\A_{-4}\B_{0}\right] |\phi_q\R\eq$$ so
that
$$\B_{-3}\A_{-1}|\phi_q\R =\left[
\A_{-1}\B_{-3}+
(c'_1-1)\A_{-2}\B_{-2}+(c'_2-c'_1)\A_{-3}\B_{-1} +(c'_3-c'_2)\A_{-4}\B_{0}\right]
|\phi_q\R\eq$$
and this is indeed a linear combination of all the independent states identified in
lemma 3.

Consider finally $\A_{-1}\A_{-2}\B_0|\phi_q\R$.  Observe that $\A_p\B_0|\phi_q\R=0$
if $p>0$.  Indeed, it can be written as the $l=0$ term of the sum $\sum_{l\geq 0}
c'_l\A_{n-1-l}\B_{m+1+l}|\phi_q\R $  with $n=p+1$ and $m=-1$; this sum can be
inverted by means of (\comma) to yield
$-\sum_{l\geq 0}
c'_l\B_{-1-l}\A_{p+1+l}|\phi_q\R$, and all these terms vanish, plus a term
proportional to $L_p|\phi_q\R$, which is also zero. With this information, we can
write
$$\eqalign{
 \A_{-1}\A_{-2}\B_0|\phi_q\R &= \left[\sum_{l=0}^\y
c_l\A_{-1-l}\A_{-2+l} - c_1\A_{-2}\A_{-1}-c_2\A_{-3}\A_{0}\right] \B_0|\phi_q\R\cr
&= \left[\sum_{l=0}^\y
c_l\A_{-2-l}\A_{-1+l} - c_1\A_{-2}\A_{-1}-c_2\A_{-3}\A_{0}\right] \B_0|\phi_q\R\cr
&= \left[(1- c_1)\A_{-2}\A_{-1}-(c_1-c_2)\A_{-3}\A_{0}\right] \B_0|\phi_q\R\cr}\eq$$
Finally, we have
$$\A_{0} \B_0|\phi_q\R= {N+2\over N}L_0|\phi_q\R= {N+2\over
N}h_{\phi_q} |\phi_q\R\eq$$
so that when acting on a highest-weight state, $\A_{-1}\A_{-2}\B_0$ can be expressed
as a linear combination of $\A_{-2}\A_{-1}\B_0$ and $\A_{-3}$.

We stress that the sum of the fractional parts of the modes in a mixed string does
not depend upon the relative position of the $\A$ and $\B$ operators in the string;
it depends only on their relative number.  More precisely, for the string (\mixst),
this sum is equal to
$${1\over N} (j-k)[q+(j-k)]\eq$$

Consider now the counting of states:

\n {\it Lemma} 4 -- The number of independent states of type $\A$ and lenght $j$
 of the form $\A_{-n_1} . . . \A_{-n_j}$
with $n_i \geq 1$,   $n_k \geq n_l$ if $k < l$ and $\sum
n_i = s$, is given by the number of partitions of the positive integer $s$ of
lenght $j$, denoted as $p^{[j]}(s)$.

In other words, $p^{[j]}(s)$ is the number of  partitions of $s$ with {\it exactly} $j$
parts.  For instance
 $p^{[2]}(8)=4$ since 8 can be decomposed in a partition of two terms in four ways:
$7+1=6+2=5+3=4+4$.

\n {\it Lemma} 5 -- The number of independent of states of type $\B$ and lenght $j$
 of the form $\B_{-m_1} . . . \B_{-m_j}$
with $m_i \geq 0$,   $m_k \geq m_l$ if $k < l$ and $\sum
m_i = s$, is given by the number of partitions of the positive integer $s$ with {\it at
most}
$j$ parts, denoted as $p^{(j)}(s)$.

There is an obvious relation between partitions having exactly $j$ parts and those of
having at most $j$ parts:
$$p^{(j)}(s) = \sum_{i=1}^j p^{[i]}(s)\eq$$
For instance
 $p^{(2)}(8)=5$ since 8 has precisely one partition into one part, 8 itself, that
contributes in addition to the 4 two-part partitions given above.

In fact, there is an even simpler relation between the type of partitions that enter in
the counting of the
$\A$ states and the ones used for counting the $\B$ states:
$$p^{[j]}(s) = p^{(j)}(s-j)\eq$$
For instance, $ p^{(3)}(5)= 5$ since 5 can be decomposed in 5 different ways in sums
of at most three integers: $5=4+1=3+2=3+1+1=2+2+1$, and we have $ p^{[3]}(8)=5$ since
there are 5 decompositions of 8 into 3 integers (obtained from the above ones by
adding a 1 to each partition -- after incorporating an appropriate number of
zeros --; for example, 4+1 is first rewritten as $4+1+0$ and then transformed to
$5+2+1$.

\n {\it Lemma 6} -- The number of states
$$\A_{-n_1}\A_{-n_2} .. .  \A_{-n_j}\B_{-m_1}\B_{-m_2} .. .  \B_{-m_k}
|\phi_q\R \eq$$with $$j-k=r\, \qquad\sum_{i=1}^j n_i=s_1\,, \qquad \sum_{i=1}^k m_i =
s_2\, ,\qquad  s_1+s_2= s\, ,\eq$$
with $r,s$
fixed (and $r$ supposed to be positive), is given by
$$\sum_{j=0}^s\sum_{s_1=j}^s  p^{(j)}(s_1-j)\,p^{(j-r)}(s-s_1)\eqlabel\nb$$

If $r$ is negative, the product $p^{(j)}(s_1-j)\,p^{(j-r)}(s-s_1)$ is replaced by
$p^{(j-|r|)}(s_1-j)\,p^{(j)}(s-s_1)$. However, it is not  necessary to treat both
cases separately since there is a charge reversal symmetry: the counting of states
is independent of the sign of the relative charge.

For the restricted partitions, we use the conventions
$$\eqalignD{
& p^{(0)}(n)= \delta_{n,0} &~\cr
& p^{(j)}(0)=1\quad &{\rm for}\quad j\geq 0\cr
&   p^{(j)}(n)=0 \quad &{\rm
for }\quad j<0, \forall\,  n  \cr}\eq$$ Useful expressions are:
$$\eqalign{
&p^{(1)}(n)=1\cr
&p^{(2)}(n)= \left[{n+2\over 2}\right]\cr
&p^{(n-1)}(n)=p(n)-1\cr
&p^{(j)}(n)= p(n) \quad {\rm if } \quad j\geq n\cr} \eq$$
(where, as before, the square bracket indicates the integer part).
To illustrate the above formula, we list the states for $1\leq s\leq 3$ and $r=0$:
$$\eqalignT{
& s=1: & \A_{-1}\B_0 & p^{(1)}(0)p^{(1)}(0)=1\cr
& s=2: & \A_{-1}\B_{-1} & p^{(0)}(1)p^{(1)}(1)=1\cr
&~     & \A_{-2}\B_0 & p^{(1)}(1)p^{(1)}(0)=1\cr
& ~ & \A_{-1}\A_{-1}\B_0\B_0 & p^{(2)}(0)p^{(2)}(0)=1\cr
& s=3: & \A_{-1}\B_{-2} & p^{(1)}(0)p^{(1)}(2)=1\cr
& ~ & \A_{-2}\B_{-1} & p^{(1)}(1)p^{(1)}(1)=1\cr
& ~ & \A_{-3}\B_{0} & p^{(1)}(2)p^{(1)}(0)=1\cr
& ~ & \A_{-1}\A_{-1}\B_{-1}\B_0 & p^{(2)}(0)p^{(2)}(1)=1\cr
& ~ & \A_{-2}\A_{-1}\B_0\B_0 & p^{(2)}(1)p^{(2)}(0)=1\cr
& ~ & \A_{-1}\A_{-1}\A_{-1}\B_0\B_0\B_0 \quad \quad &
p^{(3)}(0)p^{(3)}(0)=1\cr}\eq$$
Clearly, the $j=0$ term in the sum contributes only when $s=0$.   At a generic level
$s$, the contribution of the
$j=1$ term is simply
$$\sum_{s_1=1}^s  p^{(1)}(s_1-1)\,p^{(1)}(s-s_1)= s\eq$$
since $p^{(1)}(n)=1$ for all $n\geq 1$. On the other hand, the contribution of the $j=s$ term,
which forces $s_1=j$, is simply
$p^{(s)}(0)p^{(s)}(0)=1$. The sum (\nb) can thus be rewritten somewhat more simply
as (still pursuing the illustration of the $r=0$ case):
$$s+1+\sum_{j=2}^{s-1}\sum_{s_1=j}^s  p^{(j)}(s_1-j)\,p^{(j)}(s-s_1)\eqlabel\nbb$$
For $s=4$, it is simple to check that there are 12 terms (grouped by their value of
$j$ ordered as in the above expression):
$$\eqalign{ 5&+[ p^{(2)}(0)p^{(2)}(2)+ p^{(2)}(1)p^{(2)}(1)+
p^{(2)}(2)p^{(2)}(0)]\cr&+ [p^{(3)}(0)p^{(3)}(1)+ p^{(3)}(1)p^{(3)}(0)]
= 5+[2+1+2]+[1+1]=12\cr}\eq$$ For instance, the two states associated to
$p^{(2)}(2)p^{(2)}(0)= 2\cdot 1$ are $\A_{-3}\A_{-1}\B_0\B_0$ and $
\A_{-2}\A_{-2}\B_0\B_0$.

If the relative charge $2r$ is non-zero, the first level $s$ at which the sum is
nonvanishing is $s=r$ since $p^{(j-r)}(n)=0$ if $j<r$, irrespectively of the value of
$n$.  This first non-zero term is associated to a unique state since
$p^{(r)}(0)p^{(0)}(0)=1$.


\newsec{Parafermionic singular vectors}

We will now prove that every parafermionic highest-weight state $|\s_k\R$
has two (primary) holomorphic singular vectors, given by\foot{For simplicity, in
this section we stick to the $\sigma$ notation, but we focus on its holomorphic singular
vectors.}
$$|\chi_{N-k+1}\R\ = \A_{-1}^{N-k+1}|\s_k\R \, ,\qquad |\chi'_{k+1}\R\ =
\B_{0}^{k+1}|\s_k\R \eqlabel\sig$$ We thus want to prove that
$$\A_{p}|\xi\R =\B_{p+1}|\xi\R = L_{p+1}|\xi\R= 0\qquad {\rm for} \quad p\geq 0\eq$$
for
$|\xi\R= |\chi_{N-k+1}\R$ and
$|\chi'_{k+1}\R$.

Let us first consider the $|\chi\R$--type singular vector and start with  the action
of the $\A_p$ modes for $p\geq 0$. Consider first their action on $\A_{-1}|\s_k\R$:
$$\A_p\A_{-1}|\s_k\R= \sum_{l=0}^\y c'_l\A_{p-l}\A_{-1+l}|\s_k\R= \sum_{l=0}^\y
c'_l\A_{-1-l}\A_{p+l}|\s_k\R= 0\eqlabel\apa$$
In the first step we used the fact that only the $l=0$ term does
contribute to the sum (and $c'_0=1$); the second equality follows from (\commb)
and all the terms of this second sum vanish on a highest-weight state. Now suppose
that
$\A_p\A_{-1}^n|\s_k\R=0$ for a general value of
$n$, then
$$\A_p\A_{-1}^{n+1}|\s_k\R=\sum_{l=0}^\y c'_l\A_{p-l}\A_{-1+l}\A_{-1}^n|\s_k\R =
\sum_{l=0}^\y
c'_l\A_{-1-l}\A_{p+l} \A_{-1}^n|\s_k\R= 0\eqlabel\apb$$
In view of the recursive hypothesis, only the $l=0$ term does
contribute in the first sum, while in the second sum all terms vanish.  We have thus
shown that
$\A_p\A_{-1}^{n+1}|\s_k\R=0$ for all values of $n$.

Consider now
the action of $L_p$ for $p>0$ on $\A_{-1}^{n+1}|\s_k\R $.
For $n=0$, we have
$$L_p\A_{-1}|\s_k\R=\A_{-1}L_p|\s_k\R + {(p+1+k-N)\over N}\, A_{p-1} |\s_k\R\eq$$
and both terms vanish due to the highest-weight conditions. Suppose that
$L_p\A_{-1}^n|\s_k\R=0$ for a general value of $n$, then
$$L_p\A_{-1}^{n+1}|\s_k\R=\A_{-1}L_p\A_{-1}^n|\s_k\R + {(p+1+k+2n-N)\over N}\,
A_{p-1}\A_{-1}^n |\s_k\R = 0\eqlabel\virs$$
Indeed, the first term vanishes by the recursion hypothesis and the second vanishes
due to (\apb).

Finally, we consider the action of the $\B_p$ operators.  We start with
$p>1$ and leave for a  separate treatment the special case $p=1$.  We have (with
$c_0=1$)
$$\B_p\A_{-1}|\s_k\R= \sum_{l=0}^\y c_l\B_{p-l}\A_{-1+l}|\s_k\R= -\sum_{l=0}^\y
c_l\A_{-2-l}\B_{p+l+1}|\s_k\R +{N+2\over N} L_{p-1} |\s_k\R= 0\eqlabel\apc$$
Indeed, all the terms in the second sum vanish by the highest-weight condition and similarly
 $L_{p-1} |\s_k\R= 0$.
Again we suppose that $\B_p\A_{-1}^n|\s_k\R=0$ for a general value of $n$, and prove
the result for $n+1$, by standard steps:
$$\eqalign{
\B_p\A_{-1}^{n+1}|\s_k\R &=\phantom{-}\sum_{l=0}^\y
c_l\B_{p-l}\A_{-1+l}\A_{-1}^n|\s_k\R  \cr
& =-\sum_{l=0}^\y
c_l\A_{-2-l}\B_{p+l+1}\A_{-1}^n|\s_k\R+{N+2\over N} L_{p-1}\A_{-1}^n|\s_k\R=0\cr}
\eqlabel\apd$$ The vanishing of the first term in the last line follows from the
recursive hypothesis and the second one vanishes due to (\virs).

At this point, we have not found any constraint  on the value of $n$, i.e.,
$\A_{-1}^{n+1}|\s_k\R$ satisfies identically all the highest-weight conditions
considered so far.  The constraint on the value of $n$ follows from the consideration
of the action of the $\B_1$ mode.  Consider then
$$\eqalign{
\B_1\A_{-1}|\s_k\R &=\phantom{-} \sum_{l=0}^\y c_l\B_{1-l}\A_{-1+l}|\s_k\R\cr
& = -\sum_{l=0}^\y
c_l\A_{-2-l}\B_{2+l}|\s_k\R +\left[{N+2\over N} L_{0}+{1\over 2}
(-1+{k\over N})(-2+{k\over N})\right] |\s_k\R\cr
&= \phantom{-}\left[{N+2\over N}
h_{\s_k}+{1\over 2}(-1+{k\over N})(-2+{k\over N})\right] |\s_k\R \cr
&= \phantom{-}{N-k\over
N}\, |\s_k\R\cr}\eqlabel\ape$$  This vanishes only if $N=k$. More generally, consider
$$\eqalign{
\B_1\A_{-1}^{n+1}|\s_k\R &=\phantom{-}\sum_{l=0}^\y
c_l\B_{1-l}\A_{-1+l}\A_{-1}^n|\s_k\R  \cr
& =-\sum_{l=0}^\y
c_l\A_{-2-l}\B_{l+2}\A_{-1}^n|\s_k\R\cr
 & \quad +\left[{N+2\over N} L_{0}+{1\over
2}(-1+{k'\over N})(-2+{k'\over N})\right]\A_{-1}^n|\s_k\R\cr}
\eqlabel\apf$$
The different terms in the last sum vanish when acting on $|\s_k\R$ due to (\apd).
The value of $k'$ that appears in the last part refers to the charge of the state at
its right: hence,  $k'=k+2n$.  The eigenvalue of $L_0$ is easily found to be
$$L_0\A_{-1}^n|\s_k\R = \left [n-{1\over N}n(n+k)+h_{\s_k}\right]
\A_{-1}^n|\s_k\R\eq$$
(for this computation, the fractional part of the modes has to be reinserted.)
Substituting the value of
$h_{\s_k}$, we are thus left with
$$\B_1\A_{-1}^{n+1}|\s_k\R= {(n+1)(N-k-n)\over N}\A_{-1}^{n+1}|\s_k\R\eq$$
which is zero only if $n=N-k$.

The analysis of the other singular vector is similar and it will not be detailed.
Suffice to mention that
$$\B_p\B_{0}^{n+1}|\s_k\R = L_p \B_{0}^{n+1}|\s_k\R= \A_{p}\B_{0}^{n+1}|\s_k\R \,
\qquad p\geq 1\eqlabel\sgbb$$
for all values of $n$.  The constraint on $n$ follows from considering the
action of $\A_0$:
$$\A_0\B_{0}^{n+1}|\s_k\R = {(n+1)(k-n)\over N} \B_{0}^{n+1}|\s_k\R\eq$$
and this vanishes only if $n=k$.

It has already been mentioned that $|\mu_k\R= \A_{-1}^{N-k}|\s_k\R= \B_0^k|\s_k\R$.   The
missing highest-weight conditions in (\adad), namely, $$\A_p|\mu_k\R = \B_p|\mu_k\R  =
L_p|\mu_k\R =0 \quad {\rm for} \quad p\geq 1\eq$$ follow directly from (\sgbb).


\newsec{Parafermionic character formulae}

The character formula for the spin fields codes  the number of independent states,
level by level, with appropriate subtraction due to the presence of singular
vectors. In the Virasoro, as well as in the affine $\su(2)$ case,
this subtraction is actually an infinite process:
the two primary singular vectors both have two singular vectors themselves and it
turns out that they are identical; therefore, by subtracting the two primary
singular vectors and their descendents, we take out too much, that is, we eliminate twice
the secondary singular vectors.  To correct for this, we need to add once the
contribution of the secondary singular vectors.  This process repeats itself constantly and
that leads to a character formula expressed in terms of  alternate additions and
subtractions.

Exactly the same process appears here.  The two primary singular
vectors $|\chi_{N-k+1}\R$ and $|\chi'_{k+1}\R$ have themselves two singular vectors
each, and the two sets of secondary singular vectors coincide.


The explicit expression for the two sequences of singular vectors is as follows.
Type-I, which are built from $|\chi'_{k+1}\R$,  start with a rightmost $\B$ factor:
$$\eqalign{
 |\La_{k,\ell}^{\rm I(i)}\R &=
\B_0^{2\ell(N+2)+k+1}\,\A_{-1}^{(2\ell-1)(N+2)+k+1}\cdots
\A_{-1}^{(N+2)+k+1} \,\B_0^{k+1}|\s_k \R\cr
 |\La_{k,\ell}^{\rm I(ii)}\R &=
\A_{-1}^{(2\ell+1)(N+2)+k+1}\,\B_0^{2\ell(N+2)+k+1}\cdots
\A_{-1}^{(N+2)+k+1}\,
\B_0^{k+1}|\s_k\R\cr}\eqlabel\Isv$$
that is,
$$|\La_{k,\ell}^{\rm I(i)}\R =
\B_0^{2\ell(N+2)+k+1} |\La_{k,\ell-1}^{\rm I(ii)}\R \, , \qquad \qquad
|\La_{k,\ell}^{\rm I(ii)}\R=\A_{-1}^{(2\ell+1)(N+2)+k+1}  |\La_{k,\ell}^{\rm
I(i)}\R\eq$$ while Type-II start with a rightmost $\A$ factor:
$$\eqalign{
 |\La_{k,\ell}^{\rm II(i)}\R
&=\A_{-1}^{(2\ell+1)(N+2)-k-1}\,\B_0^{2\ell(N+2)-k-1}\cdots
\B_0^{2(N+2)-k-1}\,\A_{-1}^{(N+2)-k-1}|\s_k\R\cr
 |\La_{k,\ell}^{\rm II(ii)}\R
&=\B_0^{(2\ell+2)(N+2)-k-1}\,\A_{-1}^{(2\ell+1)(N+2)-k-1}\cdots
\B_0^{2(N+2)-k-1}\,\A_{-1}^{(N+2)-k-1}|\s_k\R\cr}\eqlabel\IIsv$$
that is,
$$|\La_{k,\ell}^{\rm II(i)}\R= \A_{-1}^{(2\ell+1)(N+2)-k-1}|\La_{k,\ell-1}^{\rm
II(ii)}\R\, , \qquad \qquad |\La_{k,\ell}^{\rm II(ii)}\R=
 \B_0^{(2\ell+2)(N+2)-k-1}|\La_{k,\ell}^{\rm I(i)}\R\eq$$
Notice that
 $$|\La_{k,0}^{\rm I(i)}\R= |\chi_{N-k+1}\R\,, \qquad \qquad |\La_{k,0}^{\rm
II(i)}\R= |\chi'_{k+1}\R\eq$$

For both types of  singular vectors, we have distinguished the cases (i)
where there is an odd number of groups of terms (a group refers to a power of
$\A_{-1}$ or
$\B_0$ and will often be called a {\it factor} for short), from cases
(ii) which have an even number of factors.

We first check a necessary condition for these vectors to be singular, which is that
their dimension be related to their charge in exactly the same way as for the spin field.

The total charge of these vectors is given by
$$\eqalignD{
 &|\La_{k,\ell}^{{\rm I(i)}}\R & : q= -2\ell(N+2)-k-2\cr
 &|\La_{k,\ell}^{{\rm I(ii)}}\R & : q= \phantom{-}2(\ell+1)(N+2)+k\cr
 &|\La_{k,\ell}^{{\rm II(i)}}\R & : q= \phantom{-}2(\ell+1)(N+2)-k-2\cr
 &|\La_{k,\ell}^{{\rm II(ii)}}\R & : q= -2\ell(N+2)+k\cr}\eqlabel\stre$$

In order to compute their dimension in a simple way, we first make a simple
observation. All these singular vectors have the structure
$$|\La\R= \C^{p+1}|\La'\R\eq$$
where $\C$ is either $\A_{-1}$ or $\B_0$. It is simple to check that the sum of the
fractional indices of the $p$ rightmost  $\C$--modes add up to zero; in fact,
reinserting these fractional parts, we have the following pattern, common to  all
cases:
$$|\La\R= \C_{(p+1)/N}\, \left [\C_{(p-1)/N}\C_{(p-3)/N} \cdots
\C_{-(p-3)/N}\C_{-(p-1)/N}\right]|\La'\R\eq$$
That makes manifest the fact that the sum of the modes of the operators
in the square bracket add up to zero.
As a result, the dimension of $|\La\R$  is
simply fixed by that of the state on which the $\C$--string acts
together with the mode of the leftmost $\C$ operator:
$$h_{\La}= h_{\La'}-{p+1\over N}\eq$$
Notice finally that the mode of this leftmost operator, when
multiplied by $N$, is equal to the number of $\C$ operators in the  expression for the singular
vector. In other words, we can read off the dimension of the singular vectors simply from the
total sum of the various powers of $\A_{-1}$ or $\B_0$ factors that it contains, multiplied by
$-1/N$:
$$h_{\La}= h_{\s_k}-{1\over N} [\# \A_{-1}+\# \B_0]\eqlabel\dirm$$

Let us then evaluate these numbers of modes in each singular vector.
 For type-I (i)  we have
$\ell+1$ $\B_0$--factors and
$\ell$  $\A_{-1}$--factors (since the type-I singular vectors start with
$\B_0^{k+1}$); for type-I (ii), we have $\ell+1$ factors of
$\A_{-1}$ and therefore  $\ell+1$
$\B_0$--factors.
The total number of $\A_{-1}$ and $\B_0$ operators in each case is respectively:
$$\eqalignT{
{\rm type -I ~(i)}:&\qquad
\# \A_{-1} &=r_{\ell-1}&\equiv \ell[\ell(N+2)+k+1]\cr
&\qquad
\# \B_{0}&= s_\ell &\equiv (\ell+1)[(\ell(N+2)+k+1]\cr
{\rm type -I ~(ii)}:&\qquad
\# \A_{-1}&= r_{\ell}&= (\ell+1)[(\ell+1)(N+2)+k+1]\cr
&\qquad
\# \B_{0}&= s_\ell&=(\ell+1)[(\ell(N+2)+k+1]\cr
}\eqlabel\syta$$
For type-II singular vectors, the analogous results are
$$\eqalignT{
{\rm type -II ~(i)}:&\qquad
\# \A_{-1}&= r'_{\ell} &\equiv (\ell+1))[(\ell+1)(N+2)-k-1]\cr
&\qquad
\# \B_0&= s'_{\ell-1} &\equiv \ell[(\ell+1)(N+2)-k-1]\cr
{\rm type -II ~(ii)}:&\qquad
\# \A_{-1}&= r'_{\ell} &= (\ell+1)[(\ell+1)(N+2)-k-1]\cr
&\qquad
\# \B_0&= s'_\ell &=(\ell+1)[(\ell+2)(N+2)-k-1]\cr
}\eqlabel\sytb$$

We can thus read off the dimensions of the singular vectors from (\dirm), (\syta) and
(\sytb).  In all cases, we can verify that the dimension $h$ and the charge $q$ are
related by
$$h= {q(N-q)\over 2N(N+2)}\eqlabel\dtdt$$
as it should for a highest-weight state.

But having made this observation, we have essentially established that these are
genuine singular vectors since the nontrivial part of the proof amounts to show that
$|\La_{k,\ell}^{{\rm I(i)}}\R$ and $|\La_{k,\ell}^{{\rm II(ii)}}\R$  are annihilated by
$\A_{0}$ while $|\La_{k,\ell}^{{\rm I(ii)}}\R$ and $|\La_{k,\ell}^{{\rm II(i)}}\R$
are annihilated by
$\B_1$.  But this reduces to a computation equivalent to (\apf) and that simply forces
a relation between the dimension of the vector and
its charge (both parametrized by $n$), which is precisely the relation (\dtdt) already
established.

In principle, we have only half of the explicit expressions of the singular
vectors. We claim that each singular vector has itself two primary singular vectors,
but we have only displayed one of them.  Indeed, consider for instance:
$$|\La_{k,0}^{\rm I(i)}\R = \,\B_0^{k+1}|\s_k\R \equiv |\s_{-k-2}\R\eq$$
where we have defined a spin state with a charge outside of the `Kac table'.  If we
extend directly the expression for the $|\s_k\R$ singular vectors to cases where  $k$ lies
outside the range $0\leq k\leq N-1$, we would have the following two primary singular
vectors:
$$ \A_{-1}^{N+2-(-k-2)-1} |\s_{-k-2}\R = \A_{-1}^{N+2+k+1} \,\B_0^{k+1}|\s_k\R\eq$$
and
$$\B_0^{(-k-2)+1}|\s_{-k-2}\R = \B_0^{-k-1}\,\B_0^{k+1}|\s_k\R= |\s_k\R\eq$$
In the second case we simply return to the highest-weight state of the previous layer,
while the first one yields a genuine new singular vector.  Actually, the expression of the other
primary singular vector does not have a simple factorizable expression when written from
$|\s_{-k-2}\R$. However, we argue below that the two singular vectors at one layer
-- these have the same value of $\ell$ and are of the same kind (i) or (ii), -- have common
primary singular vectors exactly as in the Virasoro case.  Granting this, we  have
then the  expression of all the singular vectors, starting from the
highest-weight state.

Given the close relation between the parafermionic theory and the $\suh(2)_N$ model, the
above expressions appear to be rather natural a posteriori.  Indeed, the $\suh(2)_N$
generators can be expressed in terms of the parafermionic field and a single boson $\vp$
(with OPE
$\vp(z)\,
\vp(w)\sim -\ln(z-w)$)  as follows:
$$\eqalign{
J^\dagger(z) &= \sqrt{N}\,\psi_1(z)\,e^{i\sqrt{2/N}\, \vp(z)} \cr
J^-(z) &= \sqrt{N}\,\psi_1^\dagger(z)\,e^{-i\sqrt{2/N}\, \vp(z)} \cr
J^0(z) &= i\sqrt{2N}\,\d\vp(z) \cr}\eq$$
The label $k$ of the spin field corresponds to the finite Dynkin label of the affine weight
$\lah$ at level $N$: $\lah= [N-k,k]$.  In the
Chevalley basis where
$J^-_0=f_1, \, J_{-1}^\dagger=f_0$ and $J_0^0=h_1/2$, with the action of
the $h_i$ being defined
as
$$h_i|\lah \R= \la_i |\lah\R\qquad{\rm with}\qquad \la_0= N-k,\, \quad \la_1=k\eq$$ the
sequence of affine singular vectors take the following compact form
$$f_{i+j}^{h_{i+j}+1}\cdots f_{i+1}^{h_{i+1}+1}f_{i}^{h_{i}+1}|\lah\R\eq$$
where
$$f_{i}^{h_{i}+1}|\lah\R \equiv f_{i}^{\la_{i}+1}|\lah\R\eq$$
starting with $i=0$ or 1 and with the addition of indices  defined modulo 2.   Since there
are no singular vectors in the $\uh(1)$ sector, the structure of these vectors is directly
transferred into the parafermionic theory, with $f_0=J_{-1}^\dagger\propto \A_{-1}$ and
$f_1= J^-_0\propto \B_0$. But that provides us with the missing piece of the previous argument: the embedding pattern of the $\su(2)_N$ theory is directly transposed into the parafermionic theory.

Unfortunately, the last step of the argument relies on the coset description and it is thus not intrinsically parafermionic.
To partly supply for this, we will exemplify below this embedding argument
 with few illustrative examples pertaining to the Ising model.

How do we evaluate the effect of these charged singular vectors, say in a module
with zero relative charge?  Clearly, we have to `fold' each singular vector by the
action of an appropriate number of $\A_{-1}$ or $\B_0$ factors to render the total
relative charge zero. Therefore, if the singular vector ends with a $\A_{-1}$
factor, we need to act on it with the appropriate number of $\B_0$  factors required
 to
make its relative charge zero; the fact that we act with $\B_0$ factors ensures that we
will get the lowest level at which a zero relative charge descendent appears.  Since the
resulting state is obtained by the actions of $\A_{-1}$ and $\B_0$ modes, this level is
simply the number of $\A_{-1}$ factors already contained in the expression of the singular
vector.  On the other hand, if the singular vector ends with a $\B_0$ factor, we
need to compensate with the action of the appropriate number of $\A_{-1}$ modes to
get the lowest zero relative charge descendent: but the total number of $\A_{-1}$
modes in the resulting state is simply the total number of $\B_0$ modes already
present in the original singular vector.

The expression for the levels at which the folding of the singular vectors does
occur is thus given by (\syta) and (\sytb).

To build up the character formula for $\s_k$, we need to subtract the singular
vectors of type (i) and add those of type (ii).  Denote by $V_{k,q}$ the free
module of relative charge $2q$ over $|\s_k\R$ and by $V_{k,q,\ell}^{\rm J}$ the free
module of relative charge $2q$ over $|\La_{k,\ell}^{\rm J}\R$.  The $|\s_k\R$
irreducible module $M_{k,q}$ takes thus the following form
$$M_{k,q}= V_{k,q} - \sum_{\ell=0}^\y [V_{k,q,\ell}^{\rm I(i)}+V_{k,q,\ell}^{\rm
II(i)}] + \sum_{\ell=0}^\y [V_{k,q,\ell}^{\rm I(ii)}+V_{k,q,\ell}^{\rm II(ii)}]
\eq$$
For free modules, the counting of states, level by level, is given by eq. (\nb) with
$r=q\geq 0$.  This is our main result.  We now make it more explicit.

The character of an uncharged ($q=0$) module for the spin state $\s_k$ is thus given by
$$\chi_{k,0}= q^{h_{\s_k}-c/24}\sum_{s=0}q^s g_{k,0}(s)\eq$$
where $g_{k,0}(s)$ corresponds to  the number of states at level $s$ and it is given
by $$g_{k,0}(s) = \sum_{\ell=0}^\y g_{k,0,\ell}(s)\eqlabel\nbi$$
with
$$g_{k,0,\ell}(s) = G_{0,0}\delta_{\ell,0} -G_{r_{\ell-1},s_\ell}
- G_{r'_{\ell},s'_{\ell-1}}+G_{r_{\ell},s_\ell} +G_{r'_{\ell},s'_\ell}
\eqlabel\nbu$$
where $$
G_{a,b}(s)\equiv\sum_{j=0}^s\sum_{s_1=j}^s
p^{(j-a)}(s_1-j)\,p^{(j-b)}(s-s_1)\eqlabel\nbb$$
with $a,b$ two non-negative integers.\foot{The $k$-dependence of $G$ is hidden in
$r_\ell, \,s_\ell$ and their prime versions.} Notice that
$$G_{a,b}(s)= 0\qquad {\rm if} \qquad s<\,{\rm max}\,(a,b)\eq$$

For instance, for the Ising model, $N=2$, with $k=0$, this reproduces the vacuum
Virasoro character (\isingca) while for $k=1$, this is the spin character also 
written in (\isingca).  Still another character formula for these two fields is given
by the usual Rocha-Caridi expression [\ref{A. Rocha-Caridi, in {\it Vertex Operators in
Mathematics and Physics}, ed. by J. Lepowsky et al, Mathematical Sciences Research
Institute Publications vol. {\bf 3},  Springer-Verlag, New-York (1985), 451.}] built
from a Virasoro module.  We stress however that the subtraction of the Virasoro singular
vectors occur at levels that do not match the levels at which parafermionic singular
vectors arise.

Let us work out the multiplicities of the first few levels of the vacuum
Ising character as expressed in the parafermionic language.  Up to level 8,
it takes the form:
$$q^{1/48}\,\chi_{1,1} = \sum_{s=0}^\infty q^s\,[G_{0,0}- G_{0,1}- G_{3,0}+ G_{5,1}+
G_{3,7}- \cdots]\eqlabel\iis$$
Let us list the number of states level by level associated to each of these
factors, or equivalently, evaluate these expressions $ G_{a,b}(s)$.
 We have
$$ \eqalignS{
& \qquad & G_{0,0}\qquad & G_{0,1}\qquad & G_{3,0}\qquad
&G_{5,1}\qquad &G_{3,7}\cr
&s=0: \qquad &~ 1  &~0 &~0 &~~0 &~~0\cr
&s=1: \qquad &~ 1 &~1 &~0  &~~0 &~~0\cr
&s=2: \qquad &~ 3 &~2 &~0 &~~0 &~~0\cr
&s=3: \qquad &~ 6 &~4 &~1 &~~0 &~~0\cr
&s=4: \qquad &~ 12 &~8 &~2 &~~0 &~~0\cr
&s=5: \qquad &~ 21 &~15 &~5 &~~1 &~~0\cr
&s=6: \qquad &~ 38 &~27 &~10 &~~2 &~~0\cr
&s=7: \qquad &~ 63 &~47 &~19 &~~5 &~~1\cr
&s=8: \qquad &~ 106 &~79 &~34 &~~10 &~~2\cr}\eq$$
The number of states, up to level 8 is then
$$\eqalign{
 1&+q(1-1)+q^2(3-2)+q^3(6-4-1)+q^4(12-8-2)\cr&+
q^5(21-15-5+1)+q^6(38-27-10+2)\cr&+q^7(63-47-19+5+1)+
q^8(106-79-34+10+2)+\cdots\cr
&= 1+q^2+q^3+2q^4+2q^5+3q^6+3q^7+5q^8+\cdots\cr}\eq$$
The $q$ expansion of the character (\iis) (as well as that of the spin field)
has been checked up to level 85.

Let us make use of this example to illustrate the embedding of singular vectors.
The first two singular vectors (the primary ones)  of the Ising vacuum are
$$ \B_0|0\R \qquad {\rm and}\qquad \A_{-1}^3|0\R\eq$$
and their descendents are respectively
$$ \A_{-1}^5\B_0|0\R \qquad {\rm and}\qquad \B_0^7\A_{-1}^3|0\R\eq$$
Now is $\A_{-1}^5\B_0|0\R$ a descendent of $\A_{-1}^3|0\R$? 
To see this, consider
the reexpression of $\A_{-1}^5\B_0|0\R$ in terms of basis states 
for which all $\A$ operators
are at the right (which differs thus from the basis considered previously). 
 The independent
states at level 5 and charge 8 with the $\A$ ordered at the 
right are $\B_0 \A_{-1}^5|0\R$ and
$\A_{-2}\A_{-1}^3|0\R$. Therefore, $\A_{-1}^5\B_0|0\R$ is 
necessarily a linear combination of
these two states.  But both states are manifestly descendents 
of $\A_{-1}^3|0\R$.  Similarly,
it is simple to see that $\B_0^7\A_{-1}^3|0\R$ is a descendent 
of $\B_0|0\R $: the list of all
states at level 3 and total charge
$-8$ with $\B$ operators ordered at the right is
$$\eqalign{ &
\A_{-1}^3\B_0^7|0\R\,,\quad
\A_{-2}\A_{-1}\B_0^6|0\R\,,\quad
\A_{-1}^2\B_{-1}\B_0^5|0\R\,,\quad
\A_{-1}\B_{-2}\B_0^4|0\R\,,\quad
\A_{-2}\B_{-1}\B_0^4|0\R\,,\cr &
\A_{-3}\B_0^5|0\R\,,\quad
\A_{-1}\B_{-1}^2\B_0^3|0\R\,,\quad
\B_{-3}\B_0^3|0\R\,,\quad
\B_{-2}\B_{-1}\B_0^2|0\R\,,\quad
\B_{-1}^3\B_0|0\R\,,\cr}\eq$$
(and indeed $G_{0,4}(s=3)=10$) and all these states have a rightmost factor
 $\B_0|0\R $ i.e., they are descendents of  $\B_0|0\R $.

For modules of relative charge $2t>0$, we simply replace all
$$G_{a,b}\quad \rw \quad  G_{a,b+t}\eq$$
and multiply the prefactor in the character expression (\nbi)
by $q^\Delta$ with $\Delta$, instead of being ${h_{\s_k}-c/24}$ is given by:
$$\Delta={h_{\vp_k^{(t)}}-c/24}-|t|\eqlabel\dde$$
(the absolute value is unnecessary but it makes the formula valid in
the case where $t$ is negative as indicated below). 
The factor $q^{-t}$ is introduced in order
to reshuffle the level counting from the state associated
to $\vp_k^{(t)}$, itself at level $t$
below the highest-weight state $|\s_k\R$.  If $2t<0$, we have the following
modifications: $G_{a,b}\,\rw\,G_{a-|t|,b}$ and
$q^{h_{\s_k}-c/24}\, \rw\,
q^{\Delta}$,
with $\Delta$ defined in (\dde).

Let us work out the fermion character of the Ising model.  The fermion is
expressed as $|\psi\R= \A_{-1}|0\R$.  Its character is computed as
follows: we list all the charge 2 descendents of the vacuum and subtract the
contribution of the charge $-4$ descendent of the singular vector $\A_{-1}|0\R$ and
the charge 4 descendent of the singular vector $\B_0|0\R$, etc.  The explicit form
of the first few charge 2 descendents of the vacuum are
$$\eqalign{
& s=1: \quad \A_{-1}|0\R\cr
& s=2: \quad  \A_{-2}|0\R, \,\A_{-1}^2\B_0|0\R\cr
& s=3: \quad \A_{-3}|0\R, \, \A_{-2}\A_{-1}\B_0|0\R, \, \A_{-1}^2\B_{-1}|0\R, \,
\A_{-1}^3\B_0^2|0\R\cr}\eq$$
However, when these states are counted from the fermion state $\A_{-1}|0\R$, the
above values of $s$ must be reduced by 1. Up to level 5, the fermion character
reads then:
$$q^{-1/2+1/48}\,\chi_{2,1} = G_{0,1}- G_{0,2}- G_{3,1}+ G_{5,2}+
- \cdots\eqlabel\iisi$$
The values of these $G_{j',j''}$ at each level are:
 $$ \eqalignS{
& \qquad & G_{0,1}\qquad & G_{0,2}\qquad & G_{3,1}\qquad
&G_{5,2}\qquad &~\cr
&s-1=0: \qquad &~ 1  &~0 &~0 &~~0 &~~\cr
&s-1=1: \qquad &~ 2 &~1 &~0  &~~0 &~~\cr
&s-1=2: \qquad &~ 4 &~2 &~1 &~~0 &~~\cr
&s-1=3: \qquad &~ 8 &~5 &~2 &~~0 &~~\cr
&s-1=4: \qquad &~ 15 &~9 &~5 &~~1 &~~\cr
&s-1=5: \qquad &~ 27 &~18 &~9 &~~2 &~~\cr
}\eq$$
The first few terms in the development of the fermion character are   then
$$\eqalign{
 1&+q(2-1)+q^2(4-2-1)+q^3(8-5-2)\cr
&+q^4(15-9-5+1)+
q^5(27-18-9+2)\cr
&= 1+q+q^2+q^3+2q^4+2q^5+\cdots\cr}\eq$$
which is to be compared with the $q$-expansion of $\chi_{2,1}$ in (\isingca).
Again this has been verified up to level 85 by computer.

The various characters for the three-state Potts model have also been extensively
checked.

Finally, we stress that the parafermionic characters just obtained provide new expressions
for the $\suh(2)$ string functions 
[\ref{V. Kac and D. Peterson, Adv. Math. {\bf 53} (1984) 125}].  Indeed, the character
of $\s_k$ with relative charge
$2\ell$, with $\ell=k-m$ is simply the normalized string function $c_m^k$ for the weight
$m\omega_1$ in the affine representation $\lah= [N-k,k]$ (see [\Nem, \GQ] and sect.
18.6 of [\DMS]).

\newsec{Conclusion}

In this work we have deepened some aspects of
the representation theory of the parafermionic conformal models. It leads in particular to
new parafermionic character formulae
 and thereby, to new expressions for $\su(2)_N$ string functions.
These new formulae may have interesting combinatorial descriptions which may give 
hints for their direct proof.

We should also point out that in [\ref{J. Lepowsky and M. Primc,  Contemporary Mathematics
{\bf 46} AMS, Providence, 1985.}\refname\LP], character formulae have also been worked out
directly from the  $Z_V$ algebra introduced in [\ref{L. Lepowsky and R.L. Wilson, Invent. Math.
{\bf 77} (1984) 199.}] and which is equivalent to the parafermionic algebra.  However, the
resulting characters are fermionic in structure  (in the terminology of [\ref{R. Kedem, T.R.
Klassen, B. M. McCoy and E. Melzer, Phys. Lett. {\bf B304} (1993) 263.}] -- i.e., these are
written in product form).  Equalling the bosonic-type formulae obtained here
and these fermionic characters provides further interesting identities whose direct
verification would remain an interesting problem.

In relation to the motivation
formulated in the introduction, it is fair to say that building the characters directly
from the parafermionic  modules has lead us to the most complicated  character
expressions that have been found so far for the parafermionic fields. Evenmore, the
Ising case does not provide notable simplifications: actually, it captures the
essential structure of the generic case. It certainly indicates that the rich
off-critical Ising results will not have a straightforward extension to the class of
parafermionic models.

On the other hand, as pointed
out also in the introduction, the techniques elaborated here are expected to be applicable
to more general parafermionic models for which there may be no alternative formulations in
terms of known conformal field theories.   Moreover, the results of this work should allow
us to reconsider, from a new point of view, the classification of the integrable
perturbations of  the parafermionic models, for instance, using the singular-vector
argument of [\ref{P. Di Francesco and P. Mathieu, Phys. Lett. {\bf B278} (1992) 79; P.
Mathieu and G. Watts, Nucl. Phys. {\bf B475} (1996) 361.}].  That will 
undoubtedly require
the extension of the present theories to the non-unitary sector. We hope to report on these
questions elsewhere.

\appendix{A}{Proofs related to the basis of independent states}


We first prove the first part of 
lemma 1, that is, that  on a generic highest-weight
state
$|\phi_q\R$, any state at fixed level $s$, built out of
$j$ operators
$\A$, can be expressed as a linear combination of terms of the form
$\A_{-n_1}\A_{-n_2} .. .  \A_{-n_j} |\phi_q\R$
with all
$n_i \geq 1 $ (and of course $\sum_i n_i=s\geq j$).  We thus want to show that, by means
of the commutation relations, we can always transform a term containing a positive index
into a combination of terms with all indices negative. Consider then, say, a term
of the type $$\A_{-n_1}\A_{n_2} .. .  \A_{-n_j} |\phi_q\R\eq$$
(we suppose in this and in the
following expressions that all the
$n_i$ are strictly positive).  Using the commutation relations (\commb), we can
rewrite it  under the
form
$$\sum_{\ell=0}^\y[-c_{\ell+1} \A_{-n_1}\A_{n_2-\ell-1}\A_{-n_3+\ell+1} .. .  \A_{-n_j}
+ c_\ell\A_{-n_1}\A_{-n_3-\ell}\A_{n_2+\ell} .. .  \A_{-n_j}]  |\phi_q\R\eq$$
(only the indices of the second and third modes from the left are affected). The effect is
to increase the indices toward the right: after a certain number of iterations, the
index of the second mode can be made strictly negative.  The resulting states either have
all their indices negative or there are  terms whose third mode has a positive index.  But
then we repeat the procedure with the second and third modes replaced by the third and the
fourth ones, and repeat again the whole process until we hit the last pair of modes; then,
because the rightmost mode acts on a highest-weight state, we can drop all the terms for
which this last mode has a positive index and we are left with
 only those terms  which all have strictly negative modes.

The second part of the lemma states that products of $\A$ modes can always be reorganized
such that $n_k \geq n_l$ if $ k < l $.  To prove this, consider a
term of the form
$\A_{-n_1}\A_{-n_2}\A_{-n_3} .. .  \A_{-n_j} |\phi_q\R$ with say $n_2<n_3$, with all other
terms ordered as expected (that is, for which $n_k \geq n_l$ if $ k < l $).  We
again use the commutation relations (\commb) to write it as
$$\sum_{\ell=0}^\y[-c_{\ell+1} \A_{-n_1}\A_{-n_2-\ell-1}\A_{-n_3+\ell+1} .. .  \A_{-n_j}
+ c_\ell\A_{-n_1}\A_{-n_3-\ell}\A_{-n_2+\ell} .. .  \A_{-n_j}]  |\phi_q\R\eq$$
Hence, we have reexpressed our term into a sum of terms for which  the second mode has a
larger absolute value (and, accordingly, the third mode has a lower absolute value).
Reapplying the argument, we can rewrite all terms into a combination of terms of the
type 
$\A_{-n_1}\A_{-n'_2}\A_{-n'_3} .. .  \A_{-n_j} |\phi_q\R$ for which the condition
$n'_2\geq n'_3$ is satisfied.  However, the ordering of the first and second modes or
the third and the fourth modes may no longer satisfy the initial ordering prescription,
but we then repeat the procedure for each pair of contiguous indices that are not in
proper order until all indices are non-decreasing toward the right.

The analysis
presented for the $\A$ strings applies directly for the $\B$ strings, the only modification
being that, in this case, the indices can also be zero.

Consider now mixed states.  We want to show that we can order separately the $\A$ strings
and the $\B$ strings but that there is no mixed ordering.
 That all positive indices can be
eliminated  is rather clear and does not depend upon the mixed nature of the string
of operators. It is also clear that we can move all the $\B$ operators to the right
and then use lemma 2 to order their indices.  However, this ordering is not
dependent upon the fact that the string of $\B$ operators acts on a highest-weight
state: as a result, we can order the $\A$ modes in the same way.  In other words, as far
as the characterization of states is concerned, that $\A$ modes act on a highest-weight
state or that a  $\B$--string is inserted in between does not matter. In that
regard, a positive mode does not annihilate the $\B$--string  but it produces states
already present, that is, already having the proper ordering.


\appendix{B}{From parafermionic to Virasoro singular vectors for the Ising and the
three-state Potts model}

We have  shown that the singular states of the highest-weight module of $|\s_k\R$
are necessarily charged.  There are obviously uncharged (relative to $|\s_k\R$)
descendents of these singular vectors.  For instance, the lowest descendent of the
vacuum ($|0\R=|\s_0\R$) singular vector
$\B_0|0\R$ is obtained by acting with $\A_{-1}$:
$$ \A_{-1}\B_0|0\R= {N+2\over N} L_{-1} |0\R\eq$$
as expected. Uncharged descendents of singular vectors at level two arise in the
module of $|\s_k\R$ for $k=1$ or $k=N-1$.  Let us work out their explicit form,
writing them with as much Virasoro modes as possible. The
 descendent of relative charge 4 and level two of $|\chi_2\R=
\B_0^2|\s_1\R$ is unique:
$$\A_{-1}^2\B_0^2|\s_1\R = -{N+2\over N^2}
[L_{-2}-(N+2)L_{-1}^2+N\A_{-2}\B_0]\,|\s_1\R\eqlabel\sitwo$$
(Uniqueness is obvious, e.g., we cannot act with say $\A_{-2}\A_0$ since $
\A_0\B_0^2|\s_1\R=0$ by (\sgbb).) On the other hand, the descendent of relative charge
$-4$ and level two of
$|\chi'_2\R= \A_{-1}^2|\s_{N-1}\R$ is
$$\B_0^2 \A_{-1}^2|\s_{N-1}\R = -{(N+2)(N+3)\over N^2}\left [L_{-2}-{N+2\over
N+3}L_{-1}^2-{N\over N+3}\A_{-2}\B_0\right] |\s_{N-1}\R\eqlabel\sitwoo$$

In a similar way, we can compute the level-three descendents of the singular vectors
$|\chi_3\R$ and $ |\chi'_3\R$, of relative charge $\pm 6$ respectively; they are
$\A_{-1}^3\B_0^3|\s_{2}\R$, which  is proportional to
$$\eqalign{
& \left [L_{-1}L_{-2} -{N+2\over 4}L_{-1}^3-{N+4\over
2(N+2)}L_{-3}-{7N+4\over 2(N+2)}\A_{-3}\B_0 \right.\cr
& \left. +{N(N+6)\over 4(N+2)}L_{-1}\A_{-2}\B_0+{N^2\over
2(N+2)}\A_{-2}\B_0L_{-1}\right] \,|\s_{2}\R\cr}\eqlabel\sithre$$
and
$\B_0^3\A_{-1}^3|\s_{N-2}\R$, proportional to
$$\eqalign{
& \left [L_{-1}L_{-2} -{N+2\over 3N+10}L_{-1}^3-{4(N^2+6N+9)\over
(3N+10)(N+2))}L_{-3} \right.\cr & \left.-{4(3N^2+16N+10)\over
(N^2-4)(3N+10)}\A_{-3}\B_0
  +{N(N^2+10N+12))\over
(3N^2+4N-20)(N+2)}L_{-1}\A_{-2}\B_0 \right.\cr
&\left.-{2N^2(2N+5)\over
(3N^2+4N-20)(N+2)}\A_{-2}\B_0L_{-1}\right]\, |\s_{N-2}\R\cr}\eqlabel\sithree$$

 Let us now look at the relation between these vectors and the usual 
Virasoro singular vectors of the Ising model. In that case $N=2$, so that $|\s_{N-1}\R$
reduces to $|\s_1\R$. We have thus two expressions (\sitwo) and (\sitwoo) for the
descendents of the parafermionic singular vector in the vacuum module:
$$\eqalign{
&[L_{-2}-4L_{-1}^2+2\A_{-2}\B_0]\,|\s_1\R =0\cr
&[5L_{-2}-4L_{-1}^2-2\A_{-2}\B_0] \,|\s_1\R=0\cr}\eq$$
Eliminating $\A_{-2}\B_0 |\s_{1}\R$ between these two equations yields
$$[L_{-2}-{4\over 3}L_{-1}^2]\,|\s_1\R =0\eq$$
which is the expected expression for the level-two singular vector of the Ising spin
field
$\phi_{1,2}$.  Consider next the level-two and level-three singular vectors of the
fermion
 field $\phi_{1,3}=\phi_{2,1}$.  Its parafermionic definition is $\A_{-1}|0\R$.  The descendents of the vacuum singular vectors must then be folded in the charge 2 sector.  The appropriate folding
of $\A_{-1}^3|0\R$ at the lowest possible level is obtained from two applications of $\B_0$:
$$\B_0 ^2\A_{-1}^3|0\R= [{5\over 2}L_{-1}^2\A_{-1} -2L_{-2}\A_{-1}-2\B_{-1}\A_{-1}^2]\,|0\R\eq$$
while the appropriate folding of $\B_0|0\R$ at level 3 is
$$\A_{-2}\A_{-1}\B_0|0\R= [-L_{-1}^2\A_{-1} +2L_{-2}\A_{-1}-\B_{-1}\A_{-1}^2]\,|0\R\eq$$
The elimination of the terms $\B_{-1}\A_{-1}^2|0\R$ leads to
$$[L_{-2}-{3\over 4}L_{-1}^2]\A_{-1}|0\R =0\eq$$ as it should, since $\A_{-1}|0\R= |\phi_{2,1}\R=|\psi\R$. To obtain the level-three singular vector, we compute the following descendents:
$$\eqalign{
&L_{-1}\A_{-2}\A_{-1}\B_0|0\R= [2L_{-3}\A_{-1}+2L_{-2}L_{-1}\A_{-1} -L_{-1}^3\A_{-1}-3\B_{-2}\A_{-1}^2-2\B_{-1}\A_{-2}\A_{-1}]\,|0\R\cr
&\A_{-3}\A_{-1}\B_0|0\R = [2L_{-3}\A_{-1} -\frac13L_{-1}^3\A_{-1}-2\B_{-2}\A_{-1}^2-\B_{-1}\A_{-2}\A_{-1}]\,|0\R\cr
&\B_{-1}\B_{0}\A_{-1}^3|0\R = [4L_{-3}\A_{-1} +4L_{-2}L_{-1}\A_{-1}-2L_{-1}^3\A_{-1}-6\B_{-2}\A_{-1}^2-\B_{-1}\A_{-2}\A_{-1}]\,|0\R\cr}
\eq$$
{}From these three relations, we can eliminate all the terms that cannot be written as
Virasoro modes acting on $\A_{-1}|0\R$: this leads to
$$[L_{-3}-2L_{-2}L_{-1}+\frac12L_{-1}^3]\A_{-1}|0\R= [L_{-3}-2L_{-2}L_{-1}+\frac12L_{-1}^3]|\psi\R\eq$$
which differs from the usual singular vector expression
$$[L_{-3}-4L_{-2}L_{-1}+\frac43L_{-1}^3]\,|\psi\R\eq$$
only by a term proportional to the descendent of the fermionic level-two singular vector.

Let us now consider one example associated to the three-state Potts model, for which
$N=3$. The lowest-level Virasoro singular vector is that of the parafermion $\psi_1$
which corresponds to the Virasoro field $\phi_{1,3}$. Its parafermionic description is
$\A_{-1}|0\R$.  The singular vectors that need to be considered are thus those of the
vacuum: $\B_0|0\R$ and $\A_{-1}^4|0\R$. The appropriate descendents are:
$$\A_{-2}\A_{-2}\B_{0}|0\R\,, \quad  \A_{-3}\A_{-2}\B_{0}|0\R\,, \quad
\B_0^3\A_{-1}^4|0\R \eq$$
and these generates  linear combinations of the form:
$$\left[a_1L_{-3}+a_2L_{-1}^3+a_3\B_{-1}\A_{-2}+a_4\B_{-2}\A_{-1}\right]
\A_{-1}|0\R\eq$$
The elimination of the two terms with parafermionic modes yields the correct singular vector
$$[L_{-3}-{3\over 4}L_{-1}L_{-2}+{9\over 40}L_{-1}^3]\,|\psi_1\R\eq$$

Clearly, the spin-field singular vectors of the three-state Potts model can be
obtained by the same procedure.


\appendix{C}{Recovering the free-fermion limit}

The whole structure of the parafermionic algebra and
the corresponding highest-weight modules do not manifestly degenerate  into a theory of free
fermions as $N=2$.  It is thus of interest to work out explicitly the free-fermionic
limit of some of the key results obtained in a general context. At first, we will show
how the parafermionic conformal algebra reduces to that of a free fermion and codes the
expression of the energy-momentum tensor.

When $N=2$, $\psi_1=\psi_1^\dagger$, i.e., there is a reality condition.  Hence there
is no concept of charge.  More precisely, the charge of a state boils down to its
sector, Neveu-Schwartz (NS) or Ramond (R). Therefore, we should not distinguish the
$\A$ modes from the $\B$ ones and we will denote them collectively as $b$, with
half-integer (integer) modes in the NS (R) sector respectively. Consider first
(\commb). When $N=2$, the coefficient $c'_\ell$ becomes:
$$c'_\ell = {\Gamma(\ell-1)\over \ell!\Gamma(-1)}\eq$$
which implies that
$$c'_0=1\,,\qquad c'_1= -1\,,\qquad c'_{\ell>1}=0\eq$$
Consider for instance the case $q=0$: the relation (\commb) reduces to
$$\left[(b_{3/2+n}b_{1/2+m}+b_{1/2+m}b_{3/2+n})- (b_{3/2+m}b_{1/2+n}+b_{1/2+n}b_{3/2+m})\right]|\phi_0\R\eq$$
Similar conditions are obtained in the
 $q=1$ sector. This shows that the anticommutator $\{b_{p},b_{r}\}$ is necessarily a function of the sum of the indices $p+r$.  To fix this function, we need to consider the other commutation
relation (\comma).

We first observe that when $N=2$
$$c_\ell = \ell+1\eq$$
Let the two sides of (\comma) act, not on a
general state $|\phi_q\R$, but on the vacuum state $|0\R$, which is a parafermionic
highest-weight state. For $n=-1$ and $m=0$, (\comma) reduces to
$$b_{-1/2}b_{-1/2}|0\R=0\eq$$
a result that confirms the fermionic nature of the $b$ modes.
 Setting $m=-2$ and $n\geq 3$, we find that
$$b_{-3/2+n}b_{-1/2}|0\R=-b_{-1/2}b_{-3/2+n}|0\R +
\{ b_{-3/2+n},b_{-1/2}\}|0\R= 0 \eq$$
The anticommutator $\{ b_{-3/2+n},b_{-1/2}\}$ must then vanish 
when $n\geq 3$. But since the result of this anticommutator should depend only upon the
sum of the indices, we conclude that for all values of $p$ and $r$ such that $p+r>0$, 
$\{b_{p},b_{r}\}=0$.  By symmetry this must also hold when $p+r<0$.
Consider finally the case where $m=1$ and $n=-1$, which yields
$$b_{1/2}b_{-1/2}|0\R=-b_{-1/2}b_{1/2}|0\R +
\{ b_{1/2},b_{-1/2}\}|0\R= \big[L_0+{1\over 2} n(n-1)
\left|_{n=-1}\big]|0\R= |0\R\right. \eq$$
Therefore the anticommutator $\{b_p,b_r\}$ is 1 when $p+r=0$. 
We have thus recovered the  standard anticommutation  relation of the free fermion:
 $$\{b_{p},b_{r}\}=\delta_{p+r,0}\eqlabel\cdcd$$

However, (\comma) contains much more information than the mere specification of (\cdcd). In order to extract it, let us return to the case where both sides of the equation act on a generic state
$|\phi_q\R$. By treating separately the two sectors $q=0,1$, 
and by reorganizing the sum appropriately, we find that
$$L_p= \sum_{\ell\in{\bf Z}+\epsilon} (\ell+1/2)\, b_{p-\ell}b_\ell\qquad (p\not=0)\eq$$(where $\epsilon $ is $1/2$ for NS and 0 for R)
and
$$L_0= \sum_{\ell\in{\bf Z}+\epsilon>0} \ell\,: b_{p-\ell}b_\ell: + {1\over 16}\delta_R \eq$$
where $\delta_R$ is 0 (1) in the NS (R) sector. These are the expected relations.

Consider now the structure of the singular vectors.  In the free-fermion
reinterpretation, these simply turn into highest-weight conditions.  For instance, for
the vacuum vector, we have:
$$\eqalign{ &\B_0|0\R= b_{1/2}|0\R=0\cr
&  \A_{-1}^3|0\R=  b_{3/2} b_{1/2} b_{-1/2}|0\R= - b_{3/2} b_{-1/2} b_{1/2}|0\R
+ b_{3/2} |0\R=0\cr}\eq$$
Therefore, there are no singular vectors in the fermionic
Fock space and the counting of states becomes substantially 
simplified. Moreover, since
$\A=\B$, we do not have to consider separately $\A$ and
 $\B$ strings of states.  Using the
$\A$--string description, we know, from lemma 1, 
that the states can be written in the form
$$b_{-n_1}b_{-n_2}\cdots b_{-n_j}|\phi_q\R\eq$$
with the $n_i$ non-decreasing from right to left and all $n_i>0$.
But given the fermionic nature of the $b$ modes, two adjacent $n_i$ 
cannot be equal, so that we
actually have an increasing sequence of integers.  The number of states at a given level $s$ is
thus determined by the number of partitions of $s$ into distinct parts. 
 The states can further
be separated according to their statistics, that is, their fermionic or 
bosonic nature.  In
this way, we recover the characters presented in (\isingca).

\vskip0.3cm
\centerline{\bf Acknowledgment}
We would like to thank  Y. Saint-Aubin  for useful and stimulating
discussions,
B. McCoy for clarifying remarks and for  pointing out [\LP]) to us  
and L. B\'egin for his
help in  verifying the character formulae by computer. We also thank J.-F. Carrier
and B. Paquet for  exploratory calculations related to this work.

\vskip0.3cm

\centerline{\bf REFERENCES}
\immediate\closeout\refs \vskip 0.5cm
  \message{References}\input references
\vfill\eject

\end

$$\eqalign{
& \A_{-2}\A_{-2}\B_{0}|0\R= \left[{5\over 3}L_{-3}-{5\over 18}L_{-1}^3-\B_{-1}\A_{-2}-{5\over 3} \A_{-2}\A_{-1}\right]\A_{-1}\0\R\cr
& \A_{-3}\A_{-2}\B_{0}|0\R= \left[...\right]\A_{-1}\0\R\cr
&\B_0^3\A_{-1}^4|0\R=
\left[L_{-3}-{123\over 84}L_{-1}^3+\B_{-2}\A_{-1}-{6\over 7}
\B_{-1}\A_{-2}\right]\A_{-1}\0\R\cr}\eq$$

\begintable{Coset fields for the $\suh(3)$ diagonal coset at levels
(1,1;2), their fractional conformal dimensions and their corresponding fields in the
minimal model with $c=4/5$.}
\nom{tabcosetpottsT}
$$ \vbox{\halign{
#\hfil&\quad#\hfil&\quad\hfil#\hfil\cr
\tableruleD
Coset fields &$h~\mod~1 $& minimal model fields\cr
\tablerule
$\{[1,0,0],[1,0,0];[2,0,0]\}$ &0 & $\phi_{(1,1)}, \phi_{(4,1)}$\cr
$\{[1,0,0],[0,1,0];[1,1,0]\}$ &1/15 & $\phi_{(3,3)}$\cr
$\{[1,0,0],[0,0,1];[1,0,1]\}$ &1/15 & $\phi'_{(3,3)}$\cr
$\{[1,0,0],[1,0,0];[0,1,1]\}$ &2/5 & $\phi_{(2,1)}, \phi_{(3,1)}$\cr
$\{[1,0,0],[0,1,0];[0,0,2]\}$ &2/3 & $\phi_{(4,3)}$\cr
$\{[1,0,0],[0,0,1];[0,2,0]\}$ &2/3 & $\phi'_{(4,3)}$\cr
\tableruleD
}} $$
\endtable